\documentclass[12pt]{article}

\usepackage{amsfonts}

\usepackage{amssymb}

\begin{document}

\begin{titlepage}

\begin{flushright}

CERN-TH/2000-065\\ LAPTH-782/2000 \\ hep-th/0003051 

\end{flushright}

\vspace{.5cm} 

\begin{center}

{\Large\bf Conformal primaries of $OSp(8/4,\mathbb{R})$ and BPS 
states in $AdS_4$ }\\ \vfill {\large  Sergio Ferrara$^\dagger$ and 
Emery Sokatchev$^\ddagger$ }\\ \vfill  \vspace{6pt} $^\dagger$ 
CERN Theoretical Division, CH 1211 Geneva 23, Switzerland 
\\ \vspace{6pt}

$^\ddagger$ Laboratoire d'Annecy-le-Vieux de Physique 
Th\'{e}orique\footnote[1]{UMR 5108 associ{\'e}e {\`a} 
 l'Universit{\'e} de Savoie} LAPTH, Chemin
de Bellevue - BP 110 - F-74941 Annecy-le-Vieux Cedex, France

\end{center}

\vfill

\begin{center}

{\bf Abstract} 

\end{center}
{\small We derive short UIR's of the $OSp(8/4,\mathbb{R})$ 
superalgebra of 3d $N=8$ superconformal field theories by the 
requirement that the highest weight states are annihilated by a 
subset of the super-Poincar\'{e} odd generators. We then find a 
superfield realization of these BPS saturated UIR's as ``composite 
operators" of the two basic ultrashort ``supersingleton" 
multiplets. These representations are the $AdS_4$ analogue of BPS 
states preserving different fractions of supersymmetry and are 
therefore suitable to classify perturbative and non-perturbative 
excitations of M-theory compactifications.  }

\end{titlepage}

\section{Introduction}\label{s1}

Superfield representations \cite{SS} of super-Poincar\'{e} and 
superconformal algebras have been proved to be useful tools since 
the early development of supersymmetry for several reasons. 

They provide the natural framework to formulate supersymmetric 
field theories in a ``covariant fashion" and allow one, in many 
cases, to achieve a simple understanding of the softening of 
``quantum divergences". This milder quantum behaviour of 
supersymmetric field theories is at the basis of the so-called 
``non-renormalization theorems" which are one of the striking 
features of supersymmetric quantum theories \cite{FIZ}. In modern 
language, which applies to generic supersymmetric theories, these 
non-renormalization theorems are due to the fact that 
supersymmetric field theories have some ``field representations"  
that are short, namely, the component field of highest dimension 
(which is not a total derivative) lies at a lower $\theta$ level 
than what is naively expected from a generic superfield. 

Examples of such ``short" superfields already appear in $N=1$ 4d 
supersymmetry and they are called ``chiral" \cite{fwz}. In the 
case of superconformal algebras chiral primaries have a ``ring 
structure" under multiplication and their conformal dimension is 
quantized in terms of the R $U(1)$ charge. 

In $N$-extended supersymmetry in $d=4$ as well as in other 
dimensions one needs to generalize the notion of ``chiral 
superfields". The point is that the shortening is often due to an 
interplay between the conformal dimension and the (non-Abelian) 
R-symmetry quantum numbers. The latter, in $d=3$ and 6 are related 
to the Dynkin labels of the $SO(N)$ and $USp(2N)$ groups while in 
$d=4$ for $N\geq 2$, to the Dynkin labels of $SU(N)$. 

Extended superspaces, enlarged with coordinates on $G/H$ where $G$ 
is the R-symmetry of the superconformal algebra and $H$ is a 
maximal subgroup (with rank of $H$ = rank of $G$) are called 
harmonic superspaces \cite{GIKOS,hh}. They provide the suitable 
framework in which the notion of chirality is generalized to 
Grassmann analyticity \cite{GIO}. For these  ``short" superfields 
the superconformal algebra is realized in a subspace of the full 
superspace which contains a reduced number of the original 
anticommuting Grassmann variables. 

In the spirit of the AdS/CFT correspondence \cite{AGMOO} where boundary 
``conformal operators" of $CFT_d$ are mapped onto ``bulk states" 
in $AdS_{d+1}$, multiplet shortening translates into a BPS 
condition on massive (and massless) particle states in anti-de 
Sitter space (see, for instance, \cite{FZa}). 

Superconformal algebras in $d$ dimensions appear as vacuum 
symmetries of string or M-theory compactified on $AdS_{d+1}$. 
Massive BPS saturated UIR's of these algebras should therefore be 
relevant to classify solitons preserving different fractions of 
supersymmetry, as it happens in the corresponding flat space 
limit. 

The general analysis of multiplet shortening is related to 
the so-called ``unitary bounds" of UIR's of superconformal 
algebras. For the $d=4$ case the latter was obtained in the 80's 
in Ref. \cite{ff1} for $N=1$ and in Ref. \cite{dp} for 
arbitrary $N$. The relation with the multiplet shortening and the 
$AdS_5/CFT_4$ correspondence was recently spelled out in \cite{AFSZ}.

The superfield analysis in $CFT_d$ is ``dual" to the ``state" 
analysis \cite{gm,GT,FN,GW} on $AdS_{d+1}$ since the same 
superalgebra acts on these representation spaces. However, the 
superfield approach is more powerful not only because it allows 
one to treat quantum field theories but because it leads to a 
simpler classification of ``massive representations" in the 
language of composite operators. The different BPS conditions in 
$AdS_d$ are rephrased to the different Grassmann analytic 
operators (generalizations of ``chiral operators") which exist in 
extended harmonic superspace. 

The full classification of all BPS conditions was carried out for 
$d=4,6$ superconformal algebras in Refs. \cite{FS,FS1} and it is 
extended to the $d=3$ $N=8$ superconformal algebra in the present 
paper. The appropriate superconformal algebra is in this case 
$OSp(8/4,\mathbb{R})$ which is a different non-compact form of the  
superalgebra which occurs in the $(2,0)$ theory in $d=6$. The 
latter is related to M-theory on $AdS_7\times S^4$. The former is 
appropriate to the $AdS_4\times S_7$ compactification of M-theory 
and some of its representations, both massless and massive, have 
been widely considered in the literature (see, e.g., 
\cite{GW,ggg}). 

The purpose of this paper is to extend the harmonic superspace 
analysis to the $d=3$ $N=8$ case in order to obtain all BPS states 
which may occur in $AdS_4$. These are the $AdS$ analogues of the 
1/2, 1/4 and 1/8 BPS states of Poincar\'{e} supersymmetry which occur 
in the classification of extremal black holes in supergravity 
theories \cite{FK,FMG}. Therefore BPS states in $AdS_4$ 
correspond, in particular, to anti-de Sitter black holes of $N=8$ 
gauged $SO(8)$ supergravity \cite{dWN}. 

The paper is organized as follows. In Section 2 we carry out a 
general analysis of the short highest weight UIR's of 
$OSp(8/4,\mathbb{R})$. To this end we consider 
$OSp(8/4,\mathbb{R})$ as the $N=8$ 3d superconformal algebra and 
study the conditions on the HWS's which are annihilated by all the 
$S-$ (conformal supersymmetry) generators and by a fraction (1/2, 
3/8, 1/4 or 1/8) of the $Q-$ (Poincar\'e supersymmetry) ones. As a 
result we find that the Lorentz spin of these HWS's must vanish 
and that their conformal dimension should be related to their 
$SO(8)$ Dynkin labels. Such HWS's generate series of  
representations exhibiting  1/2, 3/8, 1/4 and 1/8 BPS shortening. 
The simplest multiplets of maximal shortening (1/2 BPS) are the 
two distinct ``supersingletons". In Section 3 and 4 we realize the 
$N=8$ supersingletons first as constrained superfields in ordinary 
superspace and then as Grassmann analytic superfields in harmonic 
superspace. The latter have the advantage that their analyticity 
properties are preserved by multiplication. This allows us, in 
Section 5, to construct all composite operators obtained by 
multiplying supersingleton superfields and undergoing different 
shortenings corresponding to different BPS states in the $AdS_4$ 
bulk interpretation. We show that by tensoring only one type of 
supersingletons we can only construct 1/2, 1/4 and 1/8 BPS states, 
but by mixing the two types we can reproduce the complete 
classification of short multiplets from Section 2. In this way we 
also give an indirect proof that all the representations found in 
Section 2 are indeed unitary.

\section{Short highest weight UIR's of $OSp(8/4,\mathbb{R})$}\label{s2}

In this section we shall derive the general conditions on the 
highest weight state (HWS) of a short representation of 
$OSp(8/4,\mathbb{R})$. %(we restrict ourselves to HWS's of vanishing 
%Lorentz spin). 

The superalgebra $OSp(8/4,\mathbb{R})$ is the $N=8$ superconformal 
algebra in three dimensions (only the part of the algebra relevant 
to our argument is shown): 
\begin{eqnarray}
  && \{Q^i_\alpha, Q^j_\beta\} = 2\delta^{ij} \Gamma^\mu_{\alpha\beta} 
P_\mu\;, \qquad \{S^i_\alpha, S^j_\beta\} = 2\delta^{ij} 
\Gamma^\mu_{\alpha\beta} K_\mu\;, \label{2.1.1}\\ 
  && \{Q^i_\alpha, S^j_\beta\} = \delta^{ij} M_{\alpha\beta} + 
2\epsilon_{\alpha\beta}  (T^{ij} +  \delta^{ij} D) \;,
\label{2.1.2}\\ 
  && [D,Q^i_\alpha] = {i\over 2}Q^i_\alpha\;, \qquad [D,S^i_\alpha] = 
-{i\over 2}S^i_\alpha\;,\label{2.1.3}\\ 
  && [M_{\alpha\beta}, Q^i_\gamma] = i(\epsilon_{\gamma\alpha}  
Q^i_\beta + \epsilon_{\gamma\beta}  Q^i_\alpha )\;,\qquad 
[M_{\alpha\beta}, S^i_\gamma] = i(\epsilon_{\gamma\alpha}  
S^i_\beta + \epsilon_{\gamma\beta}  S^i_\alpha )\;,\label{2.1.6}\\ 
  && [T^{ij}, Q^k_\alpha] = i(\delta^{ki} Q^j_\alpha - \delta^{kj} 
Q^i_\alpha)\;,\qquad [T^{ij}, S^k_\alpha] = i(\delta^{ki} 
S^j_\alpha - \delta^{kj} S^i_\alpha)\;,\label{2.1.4}\\ 
  && [T^{ij}, T^{kl}] = i(\delta^{ik} T^{jl} + \delta^{jl} T^{ik} 
- \delta^{jk} T^{il} - \delta^{il} T^{jk})\;. \label{2.1.5} 
\end{eqnarray}
Here we find the following generators: $Q^i_\alpha$ of $N=8$ 
Poincar\'{e} supersymmetry carrying a 3d spinor Lorentz index 
$\alpha=1,2$ and an $SO(8)$ vector \footnote{$SO(8)$ has three 
8-dimensional representations, $8_v$, $8_s$ and  $8_c$.  Since 
these three representations are related by $SO(8)$ triality, the 
choice which one to ascribe to the supersymmetry generators is 
purely conventional. In order to be consistent with the other 
$N$-extended 3d supersymmetries where the odd generators always 
belong to the vector representation, we prefer to put an $8_v$ 
index $i$ on the supercharges.} index $i=1,\ldots,8$; $S^i_\alpha$ 
of conformal supersymmetry; $P_\mu$, $\mu=0,1,2$, of translations; 
$K_\mu$ of conformal boosts; $M_{\alpha\beta} = M_{\beta\alpha}$ 
of the 3d Lorentz group $SO(2,1)\sim SL(2,\mathbb{R})$; $D$ of 
dilations; %, i.e. the $SO(1,1)\subset SO(3,2)$; 
$T^{ij}=-T^{ji}$ of $SO(8)$. 

The definition of a short representation we adopt requires that 
its HWS is annihilated by part of the Poincar\'{e} supersymmetry 
generators $Q^i_\alpha$. Since the latter are irreducible under 
the Lorentz and R symmetries, the only way to achieve shortening 
is to break one of them. Postponing the possibility 
of dealing with the Lorentz group for a future investigation, here 
we choose to break $SO(8)$ down 
to $[SO(2)]^4\sim [U(1)]^4$ and decompose the $SO(8)$ vector 
$Q^i_\alpha$ into eight independent projections carrying different 
$U(1)$ charges. The first two such projections are: 
\begin{equation}\label{2.2}
  Q^{\pm\pm}_\alpha = {1\over 2}(Q^1_\alpha \pm Q^2_\alpha)
\end{equation}
and the corresponding charge generator is $H_1 = 2iT^{12}$, so 
that 
\begin{equation}\label{2.3}
 [H_1, Q^{\pm\pm}_\alpha ] = \pm 2i Q^{\pm\pm}_\alpha \;.
\end{equation}
Note the unusual units of charge, which are spinorial rather than 
vectorial. Let us write down one of the projections of eq. 
(\ref{2.1.2}) which will be needed in what follows: 
\begin{equation}\label{2.4}
  \{Q^{++}_\alpha, S^{--}_\beta\} = {1\over 2} M_{\alpha\beta} + 
\epsilon_{\alpha\beta}  (D-{1\over 2}H_1) \; .
\end{equation}

Similarly, we introduce the second charge  
\begin{equation}\label{2.5}
  Q^{(\pm\pm)}_\alpha = {1\over 2}(Q^3_\alpha \pm Q^4_\alpha)
\end{equation}
with generator $H_2 = 2iT^{34}$, so that 
\begin{equation}\label{2.6}
 [H_2, Q^{(\pm\pm)}_\alpha ] = \pm 2i Q^{(\pm\pm)}_\alpha 
\end{equation} 
and 
\begin{equation}\label{2.7}
  \{Q^{(++)}_\alpha, S^{(--)}_\beta\} = {1\over 2} M_{\alpha\beta} + 
\epsilon_{\alpha\beta}  (D - {1\over 2}H_2) \; . 
\end{equation}

The third and fourth charges will be introduced in a different 
way. The components $\underline{i}=5,6,7,8$ of the $8_v$ of 
$SO(8)$ form an $SO(4)$ vector. Since $SO(4)\sim SU(2)\times 
SU(2)$, we can rewrite it in spinor notation with the help of the 
Pauli matrices, e.g., $Q^{\underline{i}} \ \rightarrow \ 
Q^{\underline{a}\underline{a}'}= Q^{\underline{i}} 
(\sigma^{\underline{i}})^{\underline{a}\underline{a}'}$. Doing 
this in eq. (\ref{2.1.2}) we obtain 
\begin{equation}\label{2.8}
  \{Q^{\underline{a}\underline{a}'}_\alpha, 
S^{\underline{b}\underline{b}'}_\beta\} = {1\over 2} 
\epsilon^{\underline{a}\underline{b}} 
\epsilon^{\underline{a}'\underline{b}'}  M_{\alpha\beta} -{1\over 
2} \epsilon_{\alpha\beta}  (t^{\underline{a}\underline{b}} 
\epsilon^{\underline{a}'\underline{b}'} + 
\epsilon^{\underline{a}\underline{b}} 
t^{\underline{a}'\underline{b}'} -2 
\epsilon^{\underline{a}\underline{b}} 
\epsilon^{\underline{a}'\underline{b}'} D)  
\end{equation}
where the $SU(2)$ generators $t$ commute with the supersymmetry 
ones as follows: 
\begin{equation}\label{2.9}
  [t^{\underline{a}\underline{b}} 
,Q^{\underline{c}\underline{c}'}] = i 
(\epsilon^{\underline{c}\underline{a}}Q^{\underline{b}\underline{c}'} 
+\epsilon^{\underline{c}\underline{b}}Q^{\underline{a}\underline{c}'})\;, 
\qquad   [t^{\underline{a}'\underline{b}'} , 
Q^{\underline{c}\underline{c}'}] = 
i(\epsilon^{\underline{c}'\underline{a}'}Q^{\underline{c}\underline{b}'} 
+\epsilon^{\underline{c}'\underline{b}'}Q^{\underline{c}\underline{a}'})\;. 
\end{equation}
In this notation the two remaining charges are given by
\begin{equation}\label{2.10}
  H_3 = t^{\underline{1}\underline{2}}\;, \qquad  H_4 = 
t^{\underline{1}'\underline{2}'}\;,
\end{equation}
and by denoting ${\underline{1}}\equiv {[+]}\;, \ 
{\underline{2}}\equiv {[-]}$ and ${\underline{1}'}\equiv 
{\{+\}}\;, \ {\underline{2}'}\equiv {\{-\}}$, we find
\begin{equation}\label{2.11}
  [H_3, Q^{[\pm]\{\pm\}}] = [H_4, Q^{[\pm]\{\pm\}}] = \pm i 
Q^{[\pm]\{\pm\}}\;.
\end{equation}
The two relevant projections of  eq. (\ref{2.1.2}) now are
\begin{equation}\label{2.12}
   \{Q^{[+]\{+\}}_\alpha, S^{[-]\{-\}}_\beta\} = {1\over 2} M_{\alpha\beta} + 
\epsilon_{\alpha\beta}  (D - {1\over 2}H_3  - {1\over 2}H_4) \; , 
\end{equation}
\begin{equation}\label{2.13}
   \{Q^{[+]\{-\}}_\alpha, S^{[-]\{+\}}_\beta\} = -{1\over 2} M_{\alpha\beta} - 
\epsilon_{\alpha\beta}  (D - {1\over 2}H_3  + {1\over 2}H_4) \; . 
\end{equation} 
 
Besides the four $SO(2)$ charges, the algebra of $SO(8)$ contains 
$28-4 = 24$ generators which can be arranged into 12 ``step-up" 
operators (positive roots): 
\begin{equation}\label{2.14}
 \{{\cal T}\}_+ = \left\{
  \begin{array}{l}
    T^{++(++)}\;, \ T^{++(--)}\;, \ T^{++[\pm]\{\pm\}}\;;  \\
    T^{(++)[\pm]\{\pm\}}\;; \\
    T^{[++]} \equiv T^{[+]\{+\}[+]\{-\}}\;, \ T^{\{++\}} \equiv 
    T^{[+]\{+\}[-]\{+\}} 
  \end{array}
 \right.
\end{equation}
and their complex conjugates (negative roots). Among them only 4 
(= rank of $SO(8)$) are independent (simple roots), namely, 
$T^{[++]}, \ T^{\{++\}}, \ T^{++(--)}, \ T^{(++)[-]\{-\}}$. 

Above we have given the decomposition of two of the basic 
representations of $SO(8)$ under the particular embedding of 
$[SO(2)]^4$ that we are using here. These are the $8_v$ (the 
supersymmetry generators $Q^i$) and the adjoint $28$ (the $SO(8)$ 
generators $T^{ij}$). For future reference we also give the 
decomposition of the two spinor representations, $8_s$ ($\phi^a$, 
$a=1,\ldots,8$) and $8_c$ ($\psi^{\dot a}$, $\dot a=1,\ldots,8$): 
\begin{eqnarray}
 \phi^a &\rightarrow& \phi^{+(+)[\pm]}, \ \phi^{-(-)[\pm]}, 
\ \phi^{+(-)\{\pm\}}, \ \phi^{-(+)\{\pm\}}\label{2.15'}\\ 
 \sigma^{\dot a} &\rightarrow& \sigma^{+(+)\{\pm\}}, 
\ \sigma^{-(-)\{\pm\}}, \ \sigma^{+(-)[\pm]}, \ 
\sigma^{-(+)[\pm]}\label{2.16'} 
\end{eqnarray}
This has been obtained by successive reductions: $SO(8)\ 
\rightarrow \ SO(2)\times SO(6)\sim U(1)\times SU(4) \ \rightarrow 
\ [SO(2)]^2\times SO(4)\sim [U(1)]^2\times SU(2)\times SU(2) \ 
\rightarrow \ [SO(2)]^4 \sim [U(1)]^4$. 

Now we turn to the discussion of the representations of 
$OSp(8/4,\mathbb{R})$. Let us denote a generic (quasi primary) 
superconformal field of the $OSp(8/4,\mathbb{R})$ algebra by the 
quantum numbers of its HWS: 
\begin{equation}\label{555}
 {\cal D}(\ell, J; d_1,d_2,d_3,d_4)  
\end{equation}
where $\ell$ is the conformal dimension, $J$ is the Lorentz spin 
%(in this paper $J=0$)
and $d_1,d_2,d_3,d_4$ are the Dynkin labels (see, e.g., 
\cite{FSS}) of the $SO(8)$ R symmetry. In fact, in our scheme the 
natural labels are the four charges $q_1,q_2,q_3,q_4$ (the 
eigenvalues of $H_1,\ldots,H_4$). So, we can alternatively denote 
the HWS $|\ell, J, q_i\rangle$. The Dynkin labels 
$[d_1,d_2,d_3,d_4]$ are related to the $U(1)$ charges 
$(q_1,q_2,q_3,q_4)$ as follows: 
\begin{equation}\label{172}
  d_1 = {1\over 2} (q_1-q_2)\;, \ d_2 = {1\over 2} (q_2-q_3-q_4)\;, \ 
d_3 = q_3\;, \ d_4=q_4\;. 
\end{equation}
The above relations can be most easily derived \footnote{We are 
grateful to L. Castellani for suggesting this to us.} by comparing 
the  Dynkin labels and the charges of the HWS of the following 
four irreps: $8_v: \ [1,0,0,0] \ \leftrightarrow (2,0,0,0)$,  $28: 
\ [0,1,0,0] \ \leftrightarrow (2,2,0,0)$,  $8_s: \ [0,0,1,0] \ 
\leftrightarrow (1,1,1,0)$,  $8_c: \ [0,0,0,1] \ \leftrightarrow 
(1,1,0,1)$.  Note that (\ref{172}) implies restrictions on the 
allowed values of the charges of a HWS: 
\begin{equation}\label{173}
  q_1-q_2 =2n \geq 0 \;, \quad q_2 - q_3 - q_4 =2k \geq 0 \;, 
\quad q_3 \geq 0\;, \quad q_4 \geq 0\;. 
\end{equation}

% In principle, this HWS could also carry a Lorentz spin $J$ but in 
%this paper we restrict ourselves to the case $J=0$. 
A general HWS is defined by a subset of generators of the algebra 
which annihilate it. These include all the conformal supersymmetry 
generators: 
\begin{equation}\label{2.15}
  S^i_\alpha|\ell, J, q_i\rangle = 0
\end{equation}
(and, consequently, the boosts $K_\mu$) as well as the $SO(8)$  
``step-up" operators (\ref{2.14}): 
\begin{equation}\label{2.16}
 \{{\cal T}\}_+ |\ell, J, q_i\rangle = 0\;.
\end{equation}
The second condition defines $|\ell, J, q_i\rangle$ as the HWS of 
a UIR of $SO(8)$. A similar condition ensures irreducibility under 
the Lorentz group. Further, $|\ell, J, q_i\rangle$ should be an 
eigenstate of the generators $D,\ M^2,\ H_i$ fixing its dimension 
$\ell$, spin $J$ and charges $q_i$. 

Now, what makes a multiplet ``short" is the additional requirement 
that part of the supersymmetry charges $Q^i_\alpha$ also 
annihilate the HWS. When choosing this subset of $Q$'s we have to 
make sure that it is compatible with the rest of the conditions 
and with the algebra (\ref{2.1.1})-(\ref{2.1.5}). First of all, 
these $Q$'s must anticommute among themselves, otherwise the first 
of eqs. (\ref{2.1.1}) will yield restrictions on the momentum 
$P_\mu$. Secondly, eq. (\ref{2.16}) implies that they must form a 
closed algebra (a Cauchy-Riemann structure) with all the $SO(8)$ 
step-up operators $\{{\cal T}\}_+$. It is easy to see that such a 
subset can at most involve four supercharges. In the AdS language 
such  multiplets are called 1/2 BPS ($4= {1\over 2}\;8$ generators 
annihilate the HWS). There exist two possible choices: 
\begin{eqnarray}
\mbox{type I 1/2 BPS:}  &&Q^{++}|\ell, J, q_i\rangle = 
Q^{(++)}|\ell, J, q_i\rangle = Q^{[+]\{+\}}|\ell, J, q_i\rangle =  
\nonumber\\ 
  &&  Q^{[+]\{-\}}|\ell, J, q_i\rangle = 0
\label{2.17} 
\end{eqnarray}
or  
\begin{eqnarray}
\mbox{type II 1/2 BPS:}  &&Q^{++}|\ell, J, q_i\rangle = 
Q^{(++)}|\ell, J, q_i\rangle = Q^{[+]\{+\}}|\ell, J, q_i\rangle =  
\nonumber\\ 
  && Q^{[-]\{+\}}|\ell, J, q_i\rangle = 0\;.
\label{2.17'} 
\end{eqnarray}
Finally, conditions (\ref{2.17}) or (\ref{2.17'})  should be 
consistent with (\ref{2.15}). Using the projections (\ref{2.4}), 
(\ref{2.7}), (\ref{2.12}) and (\ref{2.13}) of eq. (\ref{2.1.2}), 
we obtain the following constraint on the charges, conformal 
weight and spin of the HWS \footnote{Such relations have been 
known from the very beginning of supersymmetry, see \cite{wz}.}: 
\begin{equation}\label{2.18}
\mbox{type I 1/2 BPS:} \qquad  q_1=q_2=q_3 = 2\ell\;, \quad 
q_4=0\;, \quad J=0\;; 
\end{equation}
\begin{equation}\label{2.18''}
\mbox{type II 1/2 BPS:} \qquad  q_1=q_2=q_4 = 2\ell\;, \quad 
q_3=0\;, \quad J=0\;, 
\end{equation}
where $2\ell \equiv m$ is a non-negative integer. Computing the 
Dynkin labels from (\ref{172}), we can say that the 1/2 BPS 
multiplets above correspond to 
\begin{equation}\label{2.18'}
\mbox{type I 1/2 BPS:} \qquad    {\cal D}(m/2, 0; 0,0,m,0)\;; 
\end{equation}
\begin{equation}\label{2.18'''}
\mbox{type II 1/2 BPS:} \qquad    {\cal D}(m/2, 0; 0,0,0,m)\;. 
\end{equation}

Besides the 1/2 BPS conditions there exist weaker shortening 
conditions. Thus, we can require that a subset of only three 
supercharges annihilate the HWS. Once again, the choice must be 
consistent with condition (\ref{2.16}), and this gives only one 
possibility: 
\begin{equation}\label{2.19}
\mbox{3/8 BPS:} \qquad   Q^{++}|\ell, J, q_i\rangle = 
Q^{(++)}|\ell, J, q_i\rangle = 
 Q^{[+]\{+\}}|\ell, J, q_i\rangle = 0\;. 
\end{equation}
This is a 3/8 BPS multiplet in the AdS language. This time the 
condition on the weight, spin and charges is 
\begin{equation}\label{2.20}
  q_1=q_2 = q_3+q_4= 2\ell\;, \quad J=0\;.
\end{equation}
Denoting $q_3=m$, $q_4=n$ where $m,n$ are non-negative integers 
and computing the  Dynkin labels, we find that this type of 
multiplet corresponds to 
\begin{equation}\label{2.21}
\mbox{3/8 BPS:} \qquad     {\cal D}(1/2(m+n), 0; 0,0,m,n)\;. 
\end{equation}
  
The next step will be to take a subset of two supercharges 
compatible with (\ref{2.16}), which is 
\begin{equation}\label{2.22}
\mbox{1/4 BPS:} \qquad    Q^{++}|\ell, J, q_i\rangle = 
Q^{(++)}|\ell, J, q_i\rangle = 0\;. 
\end{equation}
This is a 1/4 BPS multiplet in the AdS language. This time the 
condition is 
\begin{equation}\label{2.23}
  q_1=q_2 = 2\ell\;, \quad J=0\;,
\end{equation}
$q_3$ and $q_4$ being only restricted by (\ref{173}). Denoting 
$q_1=q_2 = m+n+2k$, $q_3=m$, $q_4=n$ where $m,n,k$ are 
non-negative integers, we find that this type of multiplet 
corresponds to 
\begin{equation}\label{2.24}
\mbox{1/4 BPS:} \qquad      {\cal D}(1/2(m+n)+k, 0; 0,k,m,n)\;. 
\end{equation}  

Finally, the weakest shortening condition is obtained by retaining 
only one supercharge (the HWS among the eight projections of 
$Q^i$):  
\begin{equation}\label{2.25}
\mbox{1/8 BPS:} \qquad   Q^{++}|\ell, J, q_i\rangle  = 0\;. 
\end{equation}
This is a 1/8 BPS multiplet in the AdS language. The condition in 
this case is 
\begin{equation}\label{2.26}
  q_1 = 2\ell\;, \quad J=0\;,
\end{equation}
$q_2,q_3$ and $q_4$ satisfying (\ref{173}).  Denoting $q_1 = 
m+n+2k+2l$, $q_2=m+n+2k$, $q_3=m$, $q_4=n$ where $m,n,k,l$ are 
non-negative integers, we find 
\begin{equation}\label{2.27}
 \mbox{1/8 BPS:} \qquad    {\cal D}(1/2(m+n)+k+l, 0; l,k,m,n)\;.
\end{equation}   

This concludes our abstract analysis of the possible short 
representations of  $OSp(8/4,\mathbb{R})$. Note that we are not 
directly addressing the question of whether these representations 
are unitary or not. However, in the rest of the paper we shall 
show that all of them can be realized by tensoring two elementary 
building blocks, the so-called supersingleton representations. 
Since the latter are known to be UIR's  of  $OSp(8/4,\mathbb{R})$, 
this also answers the above question affirmatively.

\section{Supersingletons}\label{s3}

Let us consider the simplest $OSp(8/4,\mathbb{R})$ representations 
of the type (\ref{2.18'}) or (\ref{2.18'''}). They are obtained by 
setting $m=1$, so they correspond to ${\cal D}(1/2, 0; 0,0,1,0)$ 
or ${\cal D}(1/2, 0; 0,0,0,1)$.  Such representations are called 
``supersingletons" \cite{NS,GW}. Each of them is just a collection 
of 8 Dirac supermultiplets \cite{Fr} made up of ``Di" and ``Rac" 
singletons \cite{ff2}. We observe that in the framework of the 
AdS/CFT correspondence \cite{M} the supersingleton describes the 
microscopic degrees of freedom of an M-2 brane with the scalars 
being the coordinates transverse to the brane which are then in 
the $8_v$ of $SO(8)$. The existence of two distinct types of $N=8$ 
3d supersingletons has first been noted in Ref. \cite{GNST}.  

Our task now will be to realize the supersingleton in $N=8$ 3d 
superspace. Consider first type I. Noting that the HWS in the 
multiplet  ${\cal D}(1/2, 0; 0,0,1,0)$ has spin 0 and the Dynkin 
labels of the $8_s$ of $SO(8)$, we take a scalar superfield 
$\Phi_a(x^\mu, \theta^\alpha_i)$ carrying an external $8_s$ index 
$a$. 

The superfield $\Phi_a$ is a reducible representation of $N=8$ 
Poincar\'{e} supersymmetry. This can be seen from the fact that the 
first fermion field in its decomposition, 
\begin{equation}\label{2}
  \Phi_a(x^\mu, \theta^\alpha_i) = \phi_a(x) + \theta^\alpha_i 
\psi_{\alpha\; ia}(x) + \ldots\;, 
\end{equation}
is reducible under $SO(8)$: $\psi_{\alpha\; ia} \ \rightarrow \ 
8_v \otimes 8_s = 8_c \oplus 56_s$. The way to achieve 
irreducibility is to impose a constraint \cite{Howe} on the 
superfield which removes the $56_s$ part of $\psi_{\alpha\; ia}$: 
\begin{equation}\label{3}
\mbox{type I:}\qquad  D^i_\alpha\Phi_a = {1\over 8}\gamma^i_{a\dot 
b}\tilde\gamma^j_{\dot b c} D^j_\alpha\Phi_c\;. 
\end{equation}
Here $D^i_\alpha$ are the covariant spinor derivatives satisfying 
the supersymmetry algebra 
\begin{equation}\label{4}
  \{D^i_\alpha,D^j_\beta\}= 
2i\delta^{ij}(\Gamma^\mu)_{\alpha\beta}\partial_\mu\;.
\end{equation}
The $SO(8)$ gamma matrices $\gamma^i_{a\dot b}$ and 
$\tilde\gamma^i_{\dot a b} = (\gamma^{iT})_{\dot a b}$ satisfy the 
Clifford algebra relations
\begin{equation}\label{5}
 \gamma^i_{a\dot b}\tilde\gamma^j_{\dot b c} +  
\gamma^j_{a\dot b}\tilde\gamma^i_{\dot b c} = 2\delta^{ij} 
\delta_{ac}\;, \qquad  \tilde\gamma^i_{\dot ab}\gamma^j_{b\dot c} 
+ \tilde\gamma^j_{\dot ab}\gamma^i_{b\dot c} = 2\delta^{ij} 
\delta_{\dot a\dot c}\;.  
\end{equation}
Using (\ref{4}) one can show that the constraint (\ref{3})  
eliminates all the components of the superfield but two:
\begin{eqnarray}
  \Phi_a(x^\mu, \theta^\alpha_i)&=&\phi_a(x) + \theta^\alpha_i 
(\gamma_i)_{a\dot b}\; \psi_{\alpha\; \dot b}(x) \nonumber\\ 
  && +\theta^{\alpha}_i\theta^{\beta}_j (\gamma_{ij})_{ab}
 \; i\partial_{\alpha\beta} \phi_b  \nonumber\\ 
  && +\theta^{\alpha}_i\theta^\beta_i\theta^{\gamma}_k (\gamma_{ijk})_{a\dot b}
 \; i\partial_{(\alpha\beta} \psi_{\gamma)\dot b}  \nonumber\\ 
  && +\theta^{\alpha}_i\theta^\beta_i\theta^\gamma_k\theta^{\delta}_l 
(\gamma_{ijkl})_{ab}\; 
 \partial_{(\alpha\beta} \partial_{\gamma\delta)}  \phi_b \label{6}
\end{eqnarray}
where $\partial_{\alpha\beta}= \partial_{\beta\alpha} = 
(\Gamma^\mu)_{\alpha\beta}\partial_\mu$ and $\gamma_{ij\ldots}$ 
are the antisymmetrized products of the $SO(8)$ gamma matrices. In 
addition,  the constraint (\ref{3}) puts these fields on shell: 
\begin{equation}\label{7}
  \square\phi_a = 0\;, \qquad 
\partial^{\alpha\beta}\psi_{\beta\; \dot a} = 0\;.
\end{equation}
Thus, the content of the constrained superfield is a massless 
multiplet of Poincar\'{e} supersymmetry consisting of a scalar in the 
$8_s$ and a spinor in the $8_c$ UIR's of $SO(8)$. 
\footnote{Superfield representations of other $OSp(N/4)$ have been 
considered in the literature \cite{IS,FFre}.} 

Note that the field equations (\ref{7}) can be obtained from a 
supersymmetric action \cite{Bduff}. Consequently, the physical fields 
$\phi_a$ and $\psi_{\alpha\; \dot a}$ have canonical dimensions 
$1/2$ and $1$, respectively. This implies that the superfield 
$\Phi_a$ has dimension $1/2$, in accord with the abstract 
representation ${\cal D}(1/2, 0; 0,0,1,0)$. 

Finally, the alternative supersingleton representation of type II 
can be realized in terms of a superfield $\Sigma_{\dot a}$ 
carrying an $8_c$ external index and satisfying the constraint 
\begin{equation}\label{3'}
\mbox{type II:}\qquad  D^i_\alpha\Sigma_{\dot a} = {1\over 
8}\tilde\gamma^i_{\dot a b}\gamma^j_{b\dot c} 
D^j_\alpha\Sigma_{\dot c}\;. 
\end{equation} 
It describes a massless multiplet consisting of a scalar 
$\sigma_{\dot a}(x)$ and a spinor $\chi_{\alpha\;a}(x)$ in the 
$8_c$ and $8_s$, correspondingly.

The problem we want to address now is how to tensor 
supersingletons. Doing it directly in terms of constrained 
superfields is quite difficult. Our alternative approach consists 
in first rewriting the constraints (\ref{3}) or (\ref{3'}) as 
analyticity conditions in harmonic superspace, after which the 
tensor multiplication becomes straightforward.  
 
\section{The supersingletons as harmonic analytic superfields} 
\label{s4}

The harmonic space suitable for our purposes is given by the coset 
\footnote{A formulation of the above multiplet in harmonic 
superspace has been proposed in Ref. \cite{Howe} (see also 
\cite{ZK} and \cite{HL} for a general discussion of 
three-dimensional harmonic superspaces).  The harmonic coset used 
in \cite{Howe} is $Spin(8)/U(4)$. Although the supersingleton 
itself does indeed live on this smaller coset, the residual 
symmetry $U(4)$ will turn out too big when we start tensoring 
different realizations of the supersingleton. For this reason we 
prefer from the very beginning to use the coset (\ref{8}) with a 
minimal residual symmetry (see also \cite{AFSZ} for a discussion 
of this point).} 
\begin{equation}\label{8}
  {SO(8)\over [SO(2)]^4} \ \sim \ {Spin(8)\over [U(1)]^4}\;.
\end{equation}
This is a $28-4 = 24$-dimensional compact manifold. Instead of 
trying to introduce explicit coordinates on it, the harmonic 
method \cite{GIKOS} prescribes to use the entire matrices of the 
fundamental representation of the group to parametrize the coset. 
The complication in the case of $SO(8)$ is that one has three 
inequivalent fundamental representations, $8_s,8_c,8_v$. The 
solution to this problem has been found in Ref. \cite{GHS}. One 
introduces three sets of harmonic variables: 
\begin{equation}\label{9}
  u_a^A\;, \ w^{\dot A}_{\dot a}\;, \ v^I_i
\end{equation}
where $A$, $\dot A$ and $I$ denote the decompositions of an $8_s$, 
$8_c$ and $8_v$ index, correspondingly, into sets of four $U(1)$ 
charges, according to the coset denominator $[U(1)]^4$ (see 
Section 2 for details). Each of the $8\times 8$ real matrices 
(\ref{9}) is a matrix of the corresponding representation of 
$SO(8)\sim Spin(8)$. This implies that all of them are orthogonal 
matrices (this is a peculiarity of $SO(8)$ due to triality): 
\begin{equation}\label{10}
  u_a^A u_a^B = \delta^{AB}\;, \quad w^{\dot A}_{\dot a} w^{\dot B}_{\dot 
a}  = \delta^{\dot A\dot B}\;, \quad v^I_i v^J_i = \delta^{IJ} 
\end{equation}
(and similarly with small and capital indices interchanged). These 
matrices supply three copies of the group space (i.e., three sets 
of 28 real variables each), and we only need one to parametrize 
the coset (\ref{8}). The condition which identifies the three sets 
of harmonic variables is 
\begin{equation}\label{11}
  u_a^A (\gamma^I)_{A\dot A} w^{\dot A}_{\dot a} = v^I_i (\gamma^i)_{a\dot 
a}\;.
\end{equation}
This relation just expresses the transformation properties of the 
gamma matrices under $SO(8)$. The reader can convince him(her)self 
that the conditions (\ref{10}), (\ref{11}) leave just one set of 
28 independent parameters by taking the infinitesimal form of the 
above matrices. Note that eq. (\ref{11}) can be viewed as the 
expression of the vector harmonics in terms of the two types of 
spinor ones. Therefore we shall choose $u,w$ as our harmonic 
variables. \footnote{Although each of the three sets of harmonic 
variables depends on the same 28 parameters, we need at least two 
sets to be able to reproduce all possible representations of 
$SO(8)$.} 
 
The idea of the harmonic description of the coset (\ref{8}) is to 
consider harmonic functions defined as functions of the above sets 
of variables modulo transformations of $[U(1)]^4$. In other words, 
a harmonic function always carries a set of four $U(1)$ charges. 
These functions are then given by their ``harmonic expansions" in 
terms of all the products of harmonic variables having the same 
charges. Take, for instance,  the function 
\begin{eqnarray}
 \phi^{+(+)[+]}(u,w) &=&\phi_a u^{+(+)[+]}_a \nonumber\\
  && + \phi_{abc} u^{+(+)[+]}_a 
 u^{+(+)[+]}_b u^{-(-)[-]}_c \nonumber\\
  && + \phi_{a\dot b\dot c} u^{+(+)[+]}_a 
 w^{+(+)\{+\}}_{\dot b} w^{-(-)\{-\}}_{\dot c} + \ldots \;.  \label{13}
\end{eqnarray}
Although the harmonic function only transforms under $[U(1)]^4$, 
the coefficients in its expansion are representations of 
$SO(8)\sim Spin(8)$. Thus, a harmonic function is a collection of 
an infinite set of irreps of $SO(8)$. 

In order to make the harmonic functions irreducible we have to 
impose differential constraints on them. To this end we introduce 
harmonic derivatives (the covariant derivatives on the coset 
(\ref{8})): 
\begin{equation}\label{14}
  D^{IJ} = u^A_a (\gamma^{IJ})^{AB}{\partial\over\partial u^B_a} + 
w^{\dot A}_{\dot a} (\gamma^{IJ})^{\dot A\dot 
B}{\partial\over\partial w^{\dot B}_{\dot a}} + v^{[I}_i 
{\partial\over\partial v^{J]}_{i}}\;. 
\end{equation}
They respect the algebraic relations  (\ref{10}), (\ref{11}) among 
the harmonic variables. Moreover, these derivatives form the 
algebra of $SO(8)$ realized on the $[U(1)]^4$ projected indices 
$A,\dot A, I$ of the harmonics. Four of them just count the four 
$U(1)$ charges, i.e. the harmonic functions are their 
eigenfunctions: 
\begin{equation}\label{151}
  H_n f^{(q_1,q_2,q_3,q_4)} (u,w) = q_n f^{(q_1,q_2,q_3,q_4)} 
(u,w)\;, \quad n=1,2,3,4\;.
\end{equation}
The remaining 24 ones are the true covariant derivatives on the 
coset. In our complex $[U(1)]^4$ notation these are 
\begin{equation}\label{16}
 \{{\cal D}\}_+ = \left\{
  \begin{array}{l}
    D^{++(++)}\;, \ D^{++(--)}\;, \ D^{++[\pm]\{\pm\}}\;;  \\
    D^{(++)[\pm]\{\pm\}}\;; \\
    D^{[++]} \equiv D^{[+]\{+\}[+]\{-\}}\;, \ D^{\{++\}} \equiv 
    D^{[+]\{+\}[-]\{+\}} 
  \end{array}
 \right.
\end{equation}
and their complex conjugates. It is clear that the 12 derivatives 
(\ref{16}) correspond to the step-up operators of $SO(8)$, see 
(\ref{2.14}). Therefore we can make a harmonic function 
irreducible by demanding that all of the derivatives (\ref{16}) 
annihilate it. In other words, this differential condition reduces 
the harmonic function to a polynomial corresponding to a highest 
weight of an $SO(8)$ irrep. For example, the constraint 
\begin{equation}\label{17}
  \{{\cal D}\}_+\phi^{+(+)[+]}(u,w) = 0 \ \Rightarrow \ 
\phi^{+(+)[+]}(u,w) = \phi_a u^{+(+)[+]}_a \;,
\end{equation}
reduces the function (\ref{13}) to a $8_s$. This can easily be 
generalized to any function of the type (\ref{151}) satisfying the 
constraint
\begin{equation}\label{171}
  \{{\cal D}\}_+ f^{(q_1,q_2,q_3,q_4)} (u,w) = 0\;.
\end{equation}
This is the defining condition of the HWS of a UIR of $SO(8)$ 
given by the Dynkin labels from eq. (\ref{172}). The function 
satisfying (\ref{171}) is thus reduced to a polynomial of the 
harmonic variables: 
\begin{eqnarray}
 && f^{(q_1,q_2,q_3,q_4)} (u,w) = 
f^{(2d_1+2d_2+d_3+d_4,2d_2+d_3+d_4,d_3,d_4)} (u,w) =   
\label{174}\\ 
  &&f_{a\ldots b\ldots c\ldots \dot 
d\ldots} (u^{+(+)[+]}_a)^{d_2+d_3} (u^{+(+)[-]}_b)^{d_2} 
(u^{+(-)\{-\}}_c)^{d_1} (w^{+(+)\{+\}}_{\dot d})^{d_1+d_4} \;. 
 \nonumber
\end{eqnarray}
 
Concluding the discussion of the harmonic coset (\ref{8}) we can 
say that if one introduces complex coordinates on it, the 
conditions (\ref{171}) take the form of (covariant) analyticity 
conditions. For this reason we can call eqs. (\ref{171}) 
``harmonic analyticity" conditions.  
 
The purpose of introducing harmonic variables is to be able to 
project the supersingleton defining constraint (\ref{3}) (or 
(\ref{3'})) in an $SO(8)$ covariant way. This means to convert the 
indices $i$ and $a$ into $U(1)$ charges with the help of the 
corresponding harmonics: $D^i_\alpha \ \rightarrow \ D^I_\alpha = 
v^I_iD^i_\alpha$ and $\Phi_a \ \rightarrow \ \Phi^A = 
u^A_a\Phi_a$. Then, using the relation (\ref{11}) it is easy to 
show that, e.g., the projection $\Phi^{+(+)[+]}$ satisfies the 
following constraints: 
\begin{equation}\label{18}
  D^{++}\Phi^{+(+)[+]} = D^{(++)}\Phi^{+(+)[+]} = D^{[+]\{\pm\}}\Phi^{+(+)[+]} = 
0\;.
\end{equation}
We see that half of the spinor derivatives annihilate the 
superfield $\Phi^{+(+)[+]}$. This is the superspace realization of 
the 1/2 BPS shortening condition (\ref{2.17}). Since these spinor 
derivatives anticommute among themselves (as follows from 
(\ref{4}) after the appropriate projections), there exists a basis 
in superspace where $\Phi^{+(+)[+]}$ becomes just a function of 
half of the odd variables as well as of the harmonic variables: 
\begin{equation}\label{19}
  \mbox{type I:}\qquad \Phi^{+(+)[+]} = \Phi^{+(+)[+]}
(x_A,\theta^{++}, \theta^{(++)}, \theta^{[+]\{\pm\}}, u,w) 
\end{equation}
where
\begin{equation}\label{191}
  x_{A\alpha\beta} = x_{\alpha\beta} + 
i\theta^{++}_{(\alpha}\theta^{--}_{\beta)}  + 
i\theta^{(++)}_{(\alpha}\theta^{(--)}_{\beta)} + 
i\theta^{[+]\{+\}}_{(\alpha}\theta^{[-]\{-\}}_{\beta)} + 
i\theta^{[+]\{-\}}_{(\alpha}\theta^{[-]\{+\}}_{\beta)}\;. 
\end{equation}
We can say that $\Phi^{+(+)[+]}$ is a ``Grassmann analytic" or a 
``short" superfield.  

So far eqs. (\ref{18}) have been derived as a corollary of the 
defining constraint (\ref{3}). In order to make the latter 
equivalent to the former we have to eliminate the harmonic 
dependence in the superfield (\ref{19}). This is done by imposing 
another set of constraints, namely, the harmonic analyticity 
conditions (\ref{171}): 
\begin{equation}\label{20}
   \{{\cal D}\}_+\Phi^{+(+)[+]}
(x,\theta^{++}, \theta^{(++)}, \theta^{[+]\{\pm\}}, u,w) = 0\;. 
\end{equation}
Note that these new constraints are compatible with (\ref{18}) 
since the two sets of derivatives form a closed algebra (a 
Cauchy-Riemann structure in the terminology of Ref. \cite{Rosly}).  
It should be stressed that eq. (\ref{20}) now has implications 
other than just restricting the harmonic dependence. The reason is 
that in the superspace basis (\ref{191}) where Grassmann 
analyticity becomes manifest some of the harmonic derivatives from 
the set $\{{\cal D}\}_+$ acquire torsion terms, e.g., $D^{++(++)} 
= \partial^{++(++)}_{u,w} + 
i\theta^{++}\Gamma^\mu\theta^{(++)}\partial_\mu$, 
$D^{++[+]\{\pm\}} = 
\partial^{++[+]\{\pm\}}_{u,w} + i\theta^{++}\Gamma^\mu
\theta^{[+]\{\pm\}}\partial_\mu$, etc. This yields space-time 
derivative constraints on the components of the superfield 
$\Phi^{+(+)[+]}\;$. All this amounts to $\Phi^{+(+)[+]}$ becoming 
``ultrashort": 
\begin{eqnarray}
 \Phi^{+(+)[+]} &=&  u^{+(+)[+]}_a \phi_a(x) \nonumber\\
 &&+(\theta^{[+]\{-\}\alpha}w^{+(+)\{+\}}_{\dot a} - 
\theta^{[+]\{+\}\alpha}w^{+(+)\{-\}}_{\dot a} \nonumber\\ 
 &&\phantom{+(} - \theta^{++\alpha} w^{-(+)[+]}_{\dot a} - \theta^{(++)\alpha} 
w^{+(-)[+]}_{\dot a})\psi_{\dot a\;\alpha}(x) \nonumber \\ 
 && + \mbox{ derivative terms}\label{21}
\end{eqnarray}
where the fields are massless. In this way we recover the content 
(\ref{6}), (\ref{7}) of the ordinary constrained superfield 
describing the supersingleton multiplet. 

It is instructive to comment on the structure of the two terms in 
eq. (\ref{21}). The first one is the component at level 0 in the 
$\theta$ expansion. It is a harmonic function of the type 
(\ref{17}), i.e. a harmonic-projected $8_s$. The situation at 
level 1 is more complicated. Originally, one finds a collection of 
spinor fields with a variety of charges. In order to find out 
which one among them is the HWS of an $SO(8)$ representation, we 
have to look at the accompanying $\theta$'s. It is easy to see 
that $\theta^{[+]\{-\}}$ can serve as a starting point for 
obtaining the rest by successive applications of the harmonic 
derivatives $\{{\cal D}\}_+$ (the step-up operators of $SO(8)$): 
\begin{equation}\label{211}
\theta^{[+]\{-\}}\ \stackrel{D^{\{++\}}}{\rightarrow} \ 
\theta^{[+]\{+\}}\ \stackrel{D^{(++)[-]\{-\}}}{\rightarrow} \ 
\theta^{(++)}\ \stackrel{D^{++(--)}}{\rightarrow} \ \theta^{++}\;.
\end{equation}
At the same time, $\theta^{[+]\{-\}}$ cannot be obtained from any 
other of the projections available in the Grassmann analytic 
superspace. As a consequence, the harmonic analyticity condition 
(\ref{20}) mixes up the corresponding spinor fields (coefficients 
at level 1 in the $\theta$ expansion), with the exception of the 
one in the term 
$\theta^{[+]\{-\}\alpha}\psi^{+(+)\{+\}}_\alpha(x,u,w)$. The 
latter must satisfy the condition $\{{\cal D}\}_+ 
\psi^{+(+)\{+\}}_\alpha =0$. This means that we are dealing with 
the HWS of the representation $(1,1,0,1)\ \leftrightarrow \ 
[0,0,0,1]$, i.e. with a $8_c$. The remaining level 1 coefficients 
are related to this HWS by harmonic equations like, e.g., 
$D^{\{++\}}\psi^{+(+)\{-\}}_\alpha = \psi^{+(+)\{+\}}_\alpha$, 
etc. In other words, they correspond to different projections 
(``lower weights") of this $8_c$. 

The same argument explains why there are no new fields beyond 
level 1. Indeed, among all the level 2 $\theta$ structures we find 
two which cannot be obtained by acting with the step-up operators 
on any other structure: 
\begin{equation}\label{212}
  \theta^{[+]\{-\}\alpha}\theta^{[+]\{-\}}_\alpha A^{(1,1,-1,2)}\;, 
\qquad \theta^{[+]\{-\}\alpha} \theta^{[+]\{+\}\beta} 
B_{(\alpha\beta)}^{(1,1,-1,0)} 
\end{equation}
corresponding to a scalar and a vector fields. Now, harmonic 
analyticity again implies that these fields should be highest 
weights of $SO(8)$ irreps, but their charges do not satisfy the 
restrictions (\ref{173}). The conclusion is that there are no such 
independent fields in the expansion of the analytic superfield 
$\Phi^{+(+)[+]}$ (more precisely, $A^{(1,1,-1,2)}=0$ and 
$B_{(\alpha\beta)}^{(1,1,-1,0)} = i\partial_{\alpha\beta}\phi_a 
u_a^{+(+)[-]}$; such terms are denoted as ``derivative terms" in 
(\ref{21})). 

In conclusion we note that the alternative form of the 
supersingleton (\ref{3'}) is described by the superfield
\begin{equation}\label{212'}
\mbox{type II:}\qquad  
\Sigma^{+(+)\{+\}}(\theta^{++},\theta^{(++)},\theta^{[\pm]\{+\}}) 
\end{equation}
satisfying the same harmonic constraints (\ref{20})  but depending 
on a different set of four odd variables. Also, the charges and 
Dynkin labels of the first component are those of an $8_c$ instead 
of $8_s$. This is the superspace realization of the 1/2 BPS 
shortening condition (\ref{2.17'}). 

\section{Short multiplets as supersingleton ``composite operators"} 
\label{s5} 

In the preceding section, with the help of the harmonic variables, 
we have been able to equivalently rewrite the supersingleton as an 
ultrashort superfield satisfying both conditions of Grassmann (eq. 
(\ref{18}) or eq. (\ref{212'})) and harmonic (eq. (\ref{20})) 
analyticity. The main advantage of this new analytic form of the 
supersingleton is the possibility to tensor copies of it in a 
straightforward way and thus to obtain series of short composite 
multiplets. As we shall show in this section, this procedure 
allows us to realize all the abstract short $OSp(8/4,\mathbb{R})$ 
multiplets of Section 2. 

We observe that in the AdS/CFT correspondence the supersingleton 
multiplet describing the dynamics of many M-2 branes is endowed 
with an internal symmetry index and composite operators are 
further restricted to be singlets under the invariance group 
\cite{AOY}. 

The simplest example  of a tensor product is obtained by taking 
$p$ identical copies of type I supersingletons,  
$(\Phi^{+(+)[+]})^p$. Clearly, it satisfies the same constraints 
of Grassmann  and harmonic  analyticity. However, the latter is 
not as strong as before. The reason is that the external charges 
of the superfield have changed, and the consequences of harmonic 
analyticity strongly depend on the charges, as the argument at the 
end of the preceding section has shown. So, for generic $p\geq 4$ 
the $\theta$ expansion goes up to the maximal level 8: 
\begin{eqnarray}
  (\Phi^{+(+)[+]})^p &=& \phi^{[0,0,p,0]}  \nonumber\\
  &+& \theta^{[+]\{-\}\alpha}\psi^{[0,0,p-1,1]}_\alpha  + \ldots \nonumber\\ 
  &+& (\theta^{[+]\{-\}})^2 A^{[0,0,p-2,2]}  + \ldots \nonumber\\ 
  &+& \theta^{[+]\{-\}\alpha} \theta^{[+]\{+\}\beta} 
B_{(\alpha\beta)}^{[0,1,p-2,0]}   + \ldots \nonumber\\ 
  &+& (\theta^{[+]\{-\}})^2\theta^{[+]\{+\}\alpha}\chi^{[0,1,p-3,1]}_\alpha  
 + \ldots 
\nonumber\\ 
  &+& \theta^{[+]\{-\}\alpha} \theta^{[+]\{+\}\beta} \theta^{(++)\gamma}
\rho_{(\alpha\beta\gamma)}^{[1,0,p-2,0]}   + \ldots \nonumber\\ 
  &+& (\theta^{[+]\{-\}})^2(\theta^{[+]\{+\}})^2 C^{[0,2,p-4,0]}  
+ \ldots \nonumber\\
  &+& (\theta^{[+]\{-\}})^2\theta^{[+]\{+\}\alpha} \theta^{(++)\beta} 
D^{[1,0,p-3,1]}_{(\alpha\beta)}  + \ldots \nonumber\\ 
  &+& \theta^{[+]\{-\}\alpha} \theta^{[+]\{+\}\beta} 
\theta^{(++)\gamma} \theta^{++\delta} 
E^{[0,0,p-2,0]}_{(\alpha\beta\gamma\delta)}   + \ldots \nonumber\\  
  &+& (\theta^{[+]\{-\}})^2 (\theta^{[+]\{+\}})^2 \theta^{(++)\alpha} 
\sigma^{[1,1,p-4,0]}_{\alpha}  + \ldots \nonumber\\ 
  &+& (\theta^{[+]\{-\}})^2 \theta^{[+]\{+\}\alpha} \theta^{(++)\beta} 
\theta^{++\gamma} \omega^{[0,0,p-3,1]}_{(\alpha\beta\gamma)}  + 
\ldots \nonumber\\ 
  &+& (\theta^{[+]\{-\}})^2 (\theta^{[+]\{+\}})^2 (\theta^{(++)})^2 
F^{[2,0,p-4,0]}   + \ldots \nonumber\\ 
  &+& (\theta^{[+]\{-\}})^2 (\theta^{[+]\{+\}})^2 \theta^{(++)\alpha} \theta^{++\beta} 
G^{[0,1,p-4,0]}_{(\alpha\beta)}   + \ldots \nonumber\\ 
  &+& (\theta^{[+]\{-\}})^2 (\theta^{[+]\{+\}})^2 
(\theta^{(++)})^2 \theta^{++\alpha}\tau_\alpha^{[1,0,p-4,0]}   + 
\ldots \nonumber\\  
  &+& (\theta^{[+]\{-\}})^2 (\theta^{[+]\{+\}})^2 (\theta^{(++)})^2 
(\theta^{++})^2 H^{[0,0,p-4,0]}   + \ldots \nonumber\\ 
  &+& \mbox{ derivative terms} \label{22} 
\end{eqnarray}
Here we have shown only the leading term at each level and of each 
Lorentz structure. This is the term whose coefficient is the HWS 
of an $SO(8)$ irrep. The other terms of the same type contain 
different harmonic projections of the same component field. 
Further, instead of the charges we have directly indicated the 
corresponding Dynkin labels of each component field. Note that the 
level in the expansion also determines the conformal dimension of 
the components (given the fact that the dimension of the first 
component is $p/2$ and that of a $\theta$ is $-1/2$). 

We see that $(\Phi^{+(+)[+]})^p$ is a short superfield (it depends 
on half of the odd variables) of the type (\ref{2.18'}), but not 
an ultrashort one, unlike the supersingleton itself (the case 
$p=1$). Still, for $p=2,3$ certain terms in the expansion 
(\ref{22}) are absent if conditions (\ref{173}) are not satisfied.  
In addition, for $p=2$ one finds conservation conditions for the 
fields of spins 2, 3/2 and 1, $\partial^{\alpha\beta} 
E^{[0,0,0,0]}_{(\alpha\beta\gamma\delta)} =\partial^{\alpha\beta} 
\rho^{[1,0,0,0]}_{(\alpha\beta\gamma)} =\partial^{\alpha\beta} 
B^{[0,1,0,0]}_{(\alpha\beta)} = 0$. This is most easily seen for 
the top spin 2 which is the only $SO(8)$ singlet in the expansion 
and hence its divergence cannot be matched by any other component.

The expansion (\ref{22}) reproduces (up to triality) the content 
of the short multiplets of  $OSp(8/4,\mathbb{R})$ found in Refs. 
\cite{FN}, \cite{GW}.

Further short multiplets can be obtained by tensoring different 
analytic superfields describing the type I supersingleton. The 
point is that in Section 4 we chose a particular projection of the  
defining constraint (\ref{3}) which lead to the analytic 
superfield  $\Phi^{+(+)[+]}$. In fact, we could have done this in 
a variety of ways, each time obtaining superfields depending on 
different halves of the total number of odd variables. If we 
decide to always leave out the lowest weight $\theta^{--}$ in the 
$8_v$ formed by the $\theta$'s, we can have four (as many as the 
rank of $SO(8)$) distinct but equivalent analytic descriptions of 
the type I  supersingleton: 
\begin{eqnarray}
  &&\Phi^{+(+)[+]}
(\theta^{++}, \theta^{(++)}, \theta^{[+]\{+\}}, 
\theta^{[+]\{-\}})\;, \nonumber\\ 
  &&\Phi^{+(+)[-]} (\theta^{++}, \theta^{(++)}, 
\theta^{[-]\{+\}}, \theta^{[-]\{-\}})\;, \nonumber\\ 
  &&\Phi^{+(-)\{+\}} (\theta^{++}, 
\theta^{(--)}, \theta^{[+]\{+\}}, \theta^{[-]\{+\}})\;, 
\nonumber\\ 
  &&\Phi^{+(-)\{-\}} 
(\theta^{++}, \theta^{(--)}, \theta^{[+]\{-\}}, 
\theta^{[-]\{-\}})\;. \label{24} 
\end{eqnarray}
Then we can tensor them in the following way: 
\begin{eqnarray}
&&(\Phi^{+(+)[+]})^{p+q+r+s}(\Phi^{+(+)[-]})^{q+r+s} 
(\Phi^{+(-)\{+\}})^{r+s}(\Phi^{+(-)\{-\}})^{s}\nonumber\\ 
  &&\ =\phi^{[r+2s,q,p,r]} +\ldots \nonumber\\
  &&\ +\theta^{[+]\{-\}}_{\alpha_1} \theta^{[+]\{+\}}_{\alpha_2} 
\theta^{(++)}_{\alpha_3} \theta^{++}_{\alpha_4} 
A^{[r+2s,q,p-2,r](\alpha_1\ldots\alpha_4)} +\ldots \label{25}\\ 
  &&\ +\theta^{[+]\{-\}}_{\alpha_1} \theta^{[+]\{+\}}_{\alpha_2} 
\theta^{(++)}_{\alpha_3} \theta^{++}_{\alpha_4}  
\theta^{[-]\{+\}}_{\alpha_5} \theta^{[-]\{-\}}_{\alpha_6} 
B^{[r+2s,q-1,p,r](\alpha_1\ldots\alpha_6)} +\ldots \nonumber\\ 
  &&\ +\theta^{[+]\{-\}}_{\alpha_1} \theta^{[+]\{+\}}_{\alpha_2} 
\theta^{(++)}_{\alpha_3} \theta^{++}_{\alpha_4}  
\theta^{[-]\{+\}}_{\alpha_5} \theta^{[-]\{-\}}_{\alpha_6} 
\theta^{(--)}_{\alpha_7} 
\chi^{[r+2s-1,q,p,r](\alpha_1\ldots\alpha_7)} +\ldots \nonumber 
\end{eqnarray}
Here we have shown the first component which belongs to the 
$SO(8)$ UIR $[r+2s,q,p,r]$ and has conformal dimension $\ell = 
{1\over 2}(p+2q+3r+4s)$ (this follows from the fact that the basic 
supersingleton has dimension $1/2$). In (\ref{25}) one can also 
see the top spin of each particular series: $J_{top}=2$ if 
$q=r=s=0$, $J_{top}=3$ if $r=s=0$ or $J_{top}=7/2$ if either 
$r\neq 0$ or $s\neq 0$. The dimension of the top spin is 
$\ell[J_{top}] = {1\over 2}(p+2q+3r+4s)+J_{top}$ (since each 
$\theta$ carries dimension $-1/2$).  Note the absence of a series 
with top spin $J=5/2$: the reason is that the tensor product of 
the different realizations (\ref{24}) of the type I supersingleton 
can depend on 4, 6 or 7 $\theta$'s but not on 5. 

The above result can be summarized as follows. By considering 
composite operators made out of type I supersingletons we have 
constructed the following series of $OSp(8/4,\mathbb{R})$ UIR's 
exhibiting $1/8$, $1/4$ or $1/2$ BPS shortening: 
\begin{eqnarray}
 {1\over 8}  \mbox{ BPS:} && {\cal D}(d_1+d_2 + {1\over 
2}(d_3+d_4), 0; d_1,d_2,d_3,d_4)\;, \quad d_1-d_4 = 2s \geq 0\;;    
 \nonumber\\
 {1\over 4}  \mbox{ BPS:} && {\cal D}(d_2 + {1\over 2}d_3, 0; 
0,d_2,d_3,0)\;;  \label{1000}\\ 
 {1\over 2}  \mbox{ BPS:} && {\cal D}({1\over 2}d_3, 0; 0,0,d_3,0)\;.  
\nonumber \end{eqnarray}

We see that tensoring only one type of supersingletons cannot 
reproduce the general result of Section 2 for all possible short 
multiplets. Most notably, in (\ref{1000}) there is no 3/8 series. 
The latter can be obtained  by mixing the two types of 
supersingletons: 
\begin{equation}\label{009}
  [\Phi^{+(+)[+]}(\theta^{++},\theta^{(++)},\theta^{[+]\{\pm\}})]^{p+q}
[\Sigma^{+(+)\{+\}}(\theta^{++},\theta^{(++)},\theta^{[\pm]\{+\}})]^{q}\;, 
\end{equation}
or the same with $\Phi$ and $\Sigma$ exchanged. Counting the 
charges and the dimension, we find exact matching with the series 
(\ref{2.21}). Further, mixing two realizations of type I and one 
of type II supersingletons, we can construct the 1/4 series 
\begin{equation}\label{00777}
  [\Phi^{+(+)[+]}]^{m+k}[\Phi^{+(+)[-]}]^{k}
[\Sigma^{+(+)\{+\}}]^{n} 
\end{equation}
which corresponds to (\ref{2.24}). Finally, the full 1/8 series 
(\ref{2.27}) (i.e., without the restriction $d_1-d_4 = 2s$ in 
(\ref{1000})) can be obtained in a variety of ways.   

\section{Conclusions}

In this paper we have analyzed all short highest weight UIR's  of 
the $OSp(8/4,\mathbb{R})$ superalgebra whose HWS's are annihilated 
by part of the super-Poincar\'{e} odd generators. In the field theory 
language, highest weight reps correspond to conformal quasi 
primary superfields. Short reps correspond to superfields which do 
not depend on some of the odd coordinates, a concept generalizing 
the notion of chiral superfields of $N=1$ 4d field theories. The 
number of distinct possibilities have been shown to correspond to 
different BPS conditions on the HWS. When the algebra is 
interpreted on the $AdS_4$ bulk, for which the 3d superconformal 
field theory corresponds to the boundary M-2 brane dynamics, these 
states appear as BPS massive excitations, such as K-K states or 
AdS black holes, of M-theory on $AdS_4\times S^7$. Since in 
M-theory there is only one type of supersingleton related to the 
M-2 brane transverse coordinates \cite{Duff1}, according to our 
analysis massive states cannot be 3/8 BPS saturated, exactly as it 
happens in M-theory on $M^4\times T^7$. Indeed, the missing 
solution was also noticed in Ref. \cite{Duff2} by studying $AdS_4$ 
black holes in gauged $N=8$ supergravity. Curiously, in the 
ungauged theory, which is in some sense the flat limit of the 
former, the 3/8 BPS states are forbidden \cite{FMG} by the 
underlying $E_{7(7)}$ symmetry of $N=8$ supergravity \cite{CJ}. 

 \section*{Acknowledgements}

We would like to thank F. Delduc, M. G\"unaydin, L. Castellani, A. 
Sciarrino and P. Sorba for enlightening discussions. E.S. is 
grateful to the TH Division of CERN for its kind hospitality. The 
work of S.F. has been supported in part by the European Commission 
TMR programme ERBFMRX-CT96-0045 (Laboratori Nazionali di Frascati, 
INFN) and by DOE grant DE-FG03-91ER40662, Task C.

\end{document}